\documentclass[review]{elsarticle}

\usepackage{lineno,hyperref}
\usepackage{amsfonts}
\usepackage{amsmath}

\modulolinenumbers[5]

\begin{document}

\begin{frontmatter}

\title{Geometric modeling of aerosol surface roughness and numerical simulation of its impact on aerosol optical properties}

\author{Jianing Zhang}
\address{Interdisciplinary Research Institute, Dalian University of Technology, Panjin, 124221, China.}


\ead{fugiya@dlut.edu.cn}


\begin{abstract}
A stochastic differential equation method is introduced for geometric modeling of dust surface roughness along with discrete differential geometry technique. Optical scattering properties are computed for randomly oriented spheroidal particles with uniformly random surface roughness. Invariant imbedding T-Matrix and geometric optics method are applied to compute light scattering properties of dust particles covering from Rayleigh to geometric optics region. We simulated optical scattering properties of feldspar with these new model particles,  which shows better performance than smooth spheroids. In addition, we also introduce entropy and relative entropy as similarity measures of particle scattering properties, especially phase functions.
\end{abstract}


\end{frontmatter}

\section{Introduction}
 
Light scattering by small particles is of great importance both in scientific research\cite{bohren:1998}, and in industrial technology\cite{kok:2017}\cite{veihelmann:2006}, examples include dusts, cloud particles, ocean particles, nano metal particles et. al.. In order to characterize these particles, especially dusts, it is necessary to know how they interact with light. 

To obtain dust optical properties, light measurements and numerical simulations can be used. In terms of numerical scattering studies, perhaps the simplest geometry is isotropic homogeneous sphere, in which Lorenz-Mie theory\cite{bohren:1998} is a standard tool for analysis and still very active. Not only has qualitative insights about scattering properties been acquired with this method, the corresponding analytical solutions serve as the benchmark for testing the accuracy and convergence behaviors of various computation software packages of light scattering. Based on Lorenz-Mie theory, light scattering properties of a large class of natural or artificial small particles\cite{bohren:1998} has been modeled assuming these scatterers to be homogeneous spheres suspending in some medium.

Despite the tremendous success of sphere in modeling of dust optical properties, any extension to non-spherical geometry has been proven to be quite useful\cite{mishchenko:2000}. For example, spheroids are frequently used for modeling dust, among all non-spherical model particles. Researchers\cite{schulz:1998, mishchenko:1994, asano:1980, dubovik:2002, nousiainen:2009, nousiainen:2011} are fascinated by spheroids due to their ability in recovering light scattering properties of dust aerosols with such dissimilar appearance. Counterintuitively, other irregular model particles\cite{bi:2009, bi:2010} generated in random manners, like Gaussian random spheres\cite{muinonen:1996}, do not perform as good as agreement with laboratory measurement as spheroids in optical modeling of small natural particles. On the other hand, all real particles's surface are roughened to some degree. Compared with other factors like shapes, chemical composition, less studies\cite{ulanowski:2006, zhang:2016, collier:2016} focus on surface roughness effects on optical properties of particles. One difficulty is generation of reasonable surface roughness. Most current light scattering researchers construct particle's surface roughness texture using Gaussian random field(GRF)\cite{kahnert:2011,grynko:2016} on spheres or prisms, which are based on random Fourier series expansion on spheres or finite planes.  Another noticeable effort in modeling surface roughness is given by the random tilt model\cite{yang:1998}, in which the local surface normal at the interception of incident ray with the particle surface is randomly tilted with the Gaussian distribution. There are also some surface roughness models that fail to take the correlation of surface roughness into consideration. 

In \cite{liu:2015},  light scattering by spheres with surface roughness has been investigated using Gaussian random sphere from Rayleigh to geometric optics region. Considering the common interest in spheroids and surface roughness effects, light scattering by large spheroids with surface roughness modeled using random tilt model has been studied in \cite{kolokolova:2015} with a improved geometric optics method(IGOM)\cite{yang:1996}. Wave optics simulation\cite{zhang:2016} for non-absorptive roughened spheroids scattering was also reported.

Although considered as an important factor, current findings on surface roughness effects were controversial and even conflicting, partly due to lack of rigous roughness model that can characterize roughness effectively and filter out effects from host particles. This paper follows our previous work\cite{zhang:2016}, in which we model rough surface geometry based on the theory of Gaussian random field on Riemann manifolds. The idea of this approach is to link the physical mechanism of surface growth to geometric modeling of surface roughness. To study the physical mechanism of surface roughness, several experimental and numerical researches has been conducted for ice particles\cite{neshyba:2016}.  In this paper, we simulate light scattering by randomly rough spheroids in our new Gaussian random field model using both T-Matrix and ray tracing method. We propose the entropy as optical similarity measure between particles. We apply this new rough model particles to simulate ensemble average scattering properties and compare the results with measurement data of dust particles. 
\section{Modeling rough dust surfaces}

Natural particles vary in shapes, of which most carry obvious surface roughness. In this section, we will discuss geometric modeling of roughness on arbitrary 2D surfaces based on random field and discrete differential geometry theory. A random field $h(\mathbf x)$ in $\mathbb{R}^n$ is a function whose value are random variables for $\mathbf{x}$.  Specifically, a GRF $h(\mathbf x)$ (height function of the surface) can arise from the random superposition of harmonic waves:
\begin{equation}
h(\mathbf x) = \sum_{\mathbf k}A(|\mathbf k|)\cos(\mathbf k \cdot \mathbf x + \phi_{\mathbf k})
\end{equation}
where $A(|\mathbf k|)$ is an amplitude spectrum, sampled from some probability distribution. The random phases $\phi_{\mathbf k}$ are uniformly independently distributed within the range $[0, 2 \pi)$. The power spectrum $C(|\mathbf k|) = A^2(|\mathbf k|)$,  containing the two-point correlation information of random fields, entirely code statistical properties of $h(\mathbf x)$. 

For generating GRF, the convolution method above is limited to simple geometry like plane or sphere, since there are well defined harmonic waves(plane, cylindrical or spherical waves). In order to generate GRF on an arbitrary 2D surface, we construct random fields through stochastic differential equation(SDE) approach\cite{lindgren:2011}. 

In the following, we will briefly review stochastic surface growth theory\cite{zhao:2001, kardar:1986} on 2D plane first and derive a correlation length which is a reasonable approximation to general 2D surfaces with small scale roughness. The energy density functional of a surface can be given by
\begin{equation}
\mathcal{H} = \frac{1}{2}\nu (\nabla h)^2 +  V(h)
\end{equation}
where $\nu$ represents "surface diffusion" and $V(h)$ potential energy. The corresponding linear Langevin equation is
\begin{equation}
\frac{\partial}{\partial t} h(\mathbf{x}, t) =  \nu \nabla^2 h - a h +  \eta(\mathbf{x}, t)
\end{equation}
where $\eta$ is a Gaussian random variable, $V(h) = a h^2$  and
\begin{eqnarray}
\mathbb E[\eta(\mathbf{x}, t)] &=& 0\\
\mathbb E[\eta(\mathbf{x}, t) \eta(\mathbf{x'}, t')] &=& 2\sigma^2 \delta(\mathbf{x}-\mathbf{x'}) \delta(t-t')
\end{eqnarray}
where $\mathbb E[\cdot]$ represents expectation value, $\sigma$ is the standard deviation of the surface height function. After making a Fourier transform, the solution at time $t$ can be obtained as
\[
h(\mathbf{k}, t) = h(\mathbf{k}, 0) e^{-(\nu k^2+a)t} +\int^{t}_0 d \tau e^{-(\nu k^2+a)(t-\tau)} \eta(\mathbf{k}, \tau)
\]
If $h(\mathbf{k}, 0) = 0$, the average of surface height at time $t$ will be 0, and the covariance function of the surface height $h$ grows as
\begin{equation}
\mathbb E[h(\mathbf{k}, t)^* h(\mathbf{k}', t)] = (2\pi)^2 \delta^2(\mathbf{k} - \mathbf{k}')(1 - e^{-2(\nu k^2+a)t}) \frac{T}{\nu k^2+a}
\end{equation}
After transforming to the coordinate space, we obtain
\begin{eqnarray*}
\mathbb E[h(\mathbf{0}, t) h(\mathbf{x}, t)] = 2\pi T\int \frac{J_0(kr) (1 - e^{-2(\nu k^2+a)t} )}{\nu k^2+a}k d k
\end{eqnarray*}

\begin{figure}
\centering
\includegraphics[width=0.5\textwidth]{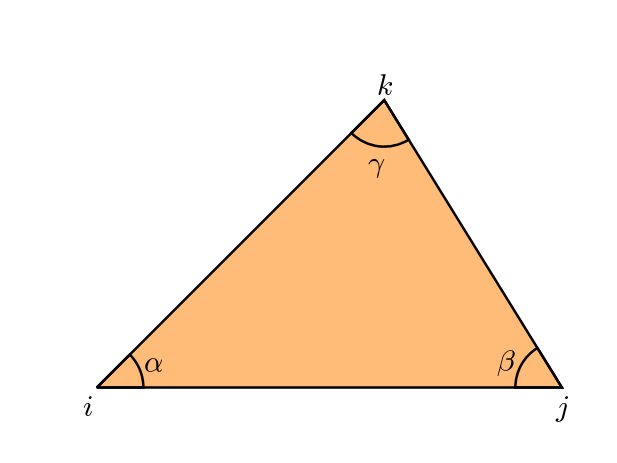}
\caption{Sample element of a typical triangular mesh.}
\label{fig:triangle}
\end{figure}

Let $\Lambda = 1/l_0$ be the cutoff. When $a \to 0$, we get

\begin{eqnarray}
\mathbb E[h(\mathbf{x}, t) h(\mathbf{x}, t)] &\propto& \int^{\Lambda}_{0} (1 - e^{-2 \nu k^2 t}) \frac{1}{k} d k\\
&=& \text{ln}[1 + 2\nu t/l^2_0]
\end{eqnarray}
where we have a correlation length $l_c = \sqrt{2\nu t}$\cite{zhao:2001}.

\[
h(\mathbf{k},\omega) = \chi(\mathbf{k},\omega)\eta(\mathbf{k},\omega)
\]
and $\chi = \frac{1}{-i\omega+\nu k^2 +a}$ is the suseptibility. According to the fluctuation-dissipation theorem, we obtain correlation function:
\[
C(\mathbf{k}, \omega) = \frac{2T}{\omega}\Im \chi(\mathbf{k}, \omega)
\] 
Now, we extend the SDE approach to generate GRFs on general 2D closed surfaces for the purpose of modeling particle surfaces. For simplicity, we adopt Edwards-Wilkinson model(a=0 in eq. 3). Following the approach given in\cite{crane:2013}, we give a review on how to approximate the differential equation on a general 2D surface using discrete differential geometry theory. Explicitly, suppose $\tilde h$ is the approximation of the true solution $h$, then:
\begin{equation}
\tilde{h} =\sum_j h_j \phi_j(\mathbf x)
\end{equation}
where $\{\phi_j\}$ is a collection of expansion functions. If we apply Green's identity to the triangle mesh, the integral is breaking up into a sum over individual triangles:
\begin{equation}
\langle \nabla \phi_i, \nabla \tilde{h} \rangle = \sum_j h_j \langle \nabla \phi_i ,\nabla \phi_j \rangle
\end{equation}
where each component is given by
\begin{equation}
\langle \nabla \phi_i, \nabla \phi_j \rangle = \int_{\Omega} dx \nabla \phi_i(x) \cdot \nabla \phi_j(x)
\end{equation}
Within a given triangle, each expansion function associated with the corresponding vertex
\begin{equation}
\langle \nabla \phi_k, \nabla \phi_k \rangle = \frac{1}{2}(\cot \alpha + \cot \beta)
\end{equation}
where $\alpha$ and $\beta$  are the remaining two angles of that given triangle (see Fig. \ref{fig:triangle}). Similarly, for the expansion functions $\phi_i$ and $\phi_j$ associated with two connected vertices along the same edge, we have
\begin{equation}
\langle \nabla \phi_i, \nabla \phi_j \rangle = -\frac{1}{2}\cot \gamma
\end{equation}
Now, we can obtain discretized version of Laplacian of arbitrary 2D surfaces, which can be approximated as
\begin{equation}
(\nabla^2 h)_{i} = \frac{1}{2} \sum_{j} (\cot \alpha_{j} + \cot \beta_{j}) (h_{j} - h_{i})
\end{equation}
where the sum is taken over the connected neighbors of vertex i.
\begin{equation}
h^{n+1}_{i} = h^{n}_{i} + \frac{1}{2} \nu \delta t \sum_{j} (\cot \alpha_{j} + \cot \beta_{j}) (h^n_{j} - h^n_{i}) + \eta_{i} \delta t
\end{equation}
Next, we describe how to generate uniform roughness on spheroid. In our simulation, dust aerosol particles are first modeled with the smooth spheroid, of which with $z$-axis as the symmetry axis the equation is given by 
\begin{equation}
\frac{x^2 + y^2}{b^2} + \frac{z^2}{a^2} = 1
\end{equation}
where $a$ is the distance from centre to either pole along the symmetry axis, and he equatorial radius of the spheroid is the semi-major axis $b$. The aspect ratio is defined as $\epsilon = a/b$. And we have
\begin{eqnarray*}
a > b : \text{prolate spheroid}\\
a < b :\text{ oblate spheroid}
\end{eqnarray*}
Then we discretize the smooth spheroid surface with triangular mesh using DistMesh\cite{persson:2004}. The normal direction at each vertex on the mesh is defined as the average of adjacent triangular facets. The surface height function $h$ is defined as the normal distance deviation from smooth surface. The simulation is performed according the above equation. The deviation and correlation of height function can be controlled by total simulation time. 
\begin{figure}
\centering
\includegraphics[width=1.0\linewidth]{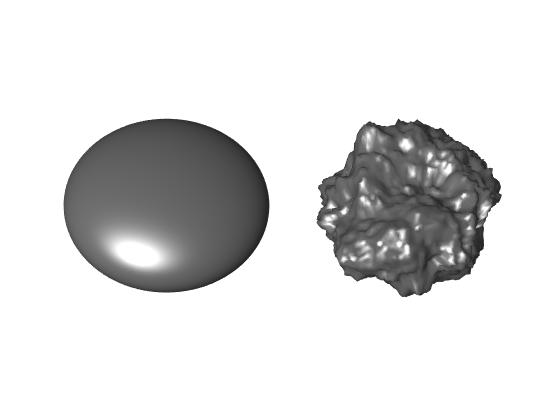}
\caption{ Comparison between smooth spheroid and rough spheroid are shown here, where $b = 1$, $ a = 0.5$. For rough sample, the variance of surface height is about 0.05 and correlation length is estimated to be 0.1.}
\label{fig:models}
\end{figure}
Fig. \ref{fig:models} illustrates a typical sample of rough spheroids. And if we keep the correlation length fixed, the standard deviation of surface height $w = \sqrt{\mathbb Var(h)}$ can be used to characterize the degree of surface roughness. The average surface area of a particle is written as 
\begin{equation}
A_r = \mathbb E[\int (1 + |\nabla h|^2)^{1/2} d\mathbf{x}]
\end{equation}
So we introduce another characteristic of roughness, which is defined as the ratio of rough and smooth surface area. In particular,  it has been noted that the above equation becomes $A_r \simeq  A_s + \frac{1}{2}\mathbb E[\int |\nabla h|^2 d\mathbf{x}]$ for small roughness $|\nabla h| \ll 1$ ($A_s$ is the area of smooth base particle.), which unveils the connection between the average surface area of rough surface and its average local slope. This roughness characteristic also connects to physical properties of natural particles. For example, the surface area of a large particle is proportional to its radiation power according to the Stefan-Boltzmann law.

\section{ Relevant light scattering theory and information entropy}
Before illustrating numerical scattering simulation with our new rough particles, we would like to discuss the relevant scattering theory first. We will suggest to introduce entropy concept to measure the similarity between particles' scattering properties. The Stokes parameters give a complete description of light fields in the sense of principle of optical equivalence. The presence of obstacles in the path of light fields produces light scattering. At the large distance, the scattered field moving in a given direction can be considered as a local plane wave field. The Stokes parameters of the incident field and scattered field are related by scattering phase matrix. Under random orientation and mirror reflection symmetry assumptions\cite{hulst:1957}, the scattering matrix is block diagonal (only 6 elements are independent)
\begin{equation}
 P(\mu)= \begin{bmatrix}
    P_{11} & P_{12} & 0
    & 0 \\
    P_{12} & P_{22} & 0
    & 0 \\
    0 & 0 & P_{33}
    & P_{34} \\
    0 & 0 & -P_{34}
    & P_{44}
  \end{bmatrix}
\end{equation}
where $\mu = \cos\theta_{sc} $, $\theta_{sc}$ is the scattering angle between incident direction and scattered direction of light.  Phase matrix $P$ are defined with the normalization condition:
\begin{equation*}
\frac{1}{2} \int P_{11}(\mu)d\mu = 1
\end{equation*}
Given the frequency of light, the phase function $P_{11}$ of a scatterer can be interpreted the angular distribution of light field intensity scattered by the object. 

\begin{figure}
\centering
\includegraphics[width=1.0\linewidth]{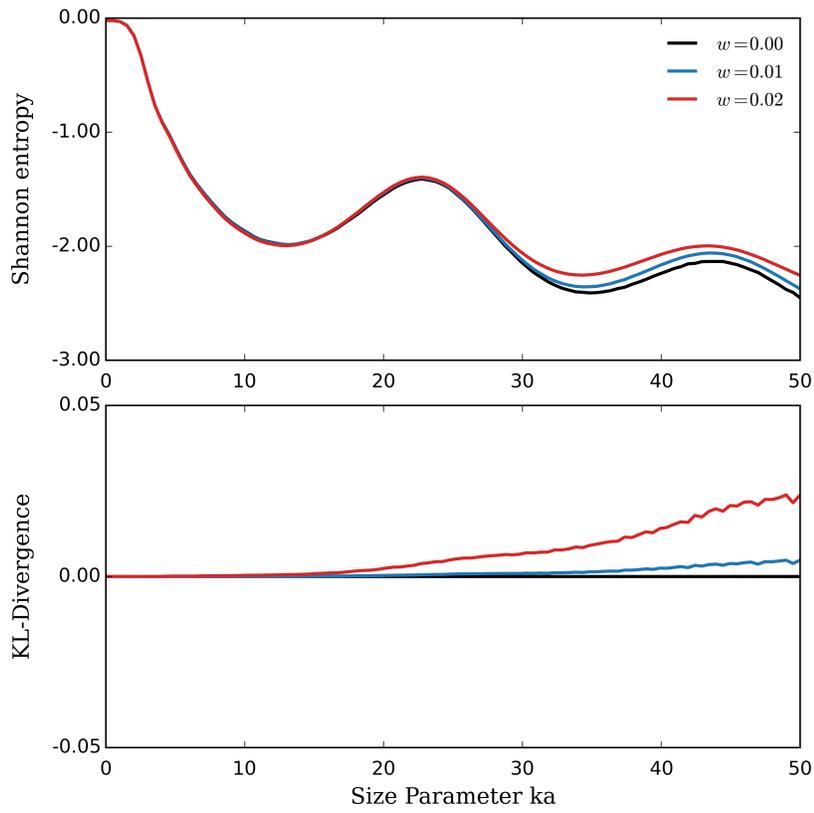}
\caption{Shannon information entropy and KL-Divergence computed by the scattering phase functions of spheroids with different degrees of surface roughness$ w = 0.00, 0.01, 0.02$. The range of the size parameter is $X \in [0, 50]$.}
\label{fig:entropy}
\end{figure}

As we had mentioned above, random irregular particles usually do not fit measured scattering phase matrix of natural particles like dust aerosols as well as spheroids. There are long and severe debates on why spheroids model optical properties of real particles so well. In this paper, we propose a potential tool to solve this debate by introducing the concept of scattering information entropy. In addition, we note that the entropy of 2D image was proposed for characterizing particle roughness in light scattering experiments\cite{ulanowski:2014}. Generally, Shannon's information entropy for any probability distribution of random variable is defined as
\begin{equation}
S=-\int d\mu P_{11}(\mu)\ln[P_{11}(\mu)]
\end{equation}
Since scattering phase function is indeed a probability density distribution of scattered directions. For each phase function, we can calculate its entropy, the degree of disorder. Overall, for larger particles, the higher peak in forward direction will correspond to lower entropy. To quantify optical similarity between different particles, we introduce the distance between the two distributions $P_{11}(\mu)$ and $P'_{11}(\mu)$ is given by the relative entropy or more formally, Kullback-Leibler divergence\cite{mackay:2003}
\begin{equation}
KL(P_{11}||P'_{11}) = \int P_{11}(\mu) \ln[P_{11}(\mu)/P'_{11}(\mu)] d\mu
\end{equation}
As an example, the scattering phase matrix of small spheroidal particles with refractive index $n =1.31$ is computed with the invariant imbedding T-matrix method (II-TM) \cite{johnson:1988, bi:2013}. Fig. \ref{fig:entropy} (top) shows the Shannon entropy as a function of size parameter $X = ka$(k is the wavenumber, and a is maximum dimension of the corresponding particle.). The pattern shows the entropy will decrease in trend but fluctuate around certain sizes. The amplitude of fluctuations decay gradually as particles' size parameter increases. It seems reasonable that entropy curves will become much smoother when particles' size go to the geometrical optics region. For $X > 20$, the curve of Shannon entropy begin to split. The surface roughness increase the Shannon entropy for each size.   
In Fig. \ref{fig:entropy} (bottom), it shows that as standard deviation $w$ (roughness metric) increases, so does the KL-divergence between rough spheroids and the corresponding smooth ones. It also indicates that as size parameter increases, the KL-divergence grows too. Therefore,  KL-divergences may help us understanding the mysterious success of spheroids, optical distance(KL-divergence) among spheroid ensemble are generally larger than random irregular particle ensemble, detailed study about this will be presented in our future work. In other words, functional spaces of phase matrix elements for spheroids span larger than that of random irregular particles. 

\section{Numerical simulation}
In the simulation, we adopt rough spheroids described in section 2 as model particles. Light scattering by particles with our new random rough surfaces are simulated using II-TM and IGOM. First, II-TM computes particles’ optical properties based on volume integral equation formalism of Maxwell’s equations\cite{johnson:1988}, in which the invariant imbedding technique is used to solve the volume integration efficiently. Compared with surface integral formalism of T-Matrix method, II-TM can deal with highly non-spherical and inhomogeneous cases. Like other T-matrix methods\cite{mishchenko:2000}, II-TM can obtain random orientation cross sections and scattering phase matrix analytically.  For large size particles, T-Matrix method falls down. Hence, IGOM which hybridizes geometric optics (ray-tracing) and physical optics methods, is utilized to perform the computation. And each ray in IGOM is generated from a uniform random incident direction, random oriented results are obtained after we perform the near field to far field mapping\cite{yang:1998}. 

\begin{figure}
\centering
\includegraphics[width=1.0\linewidth]{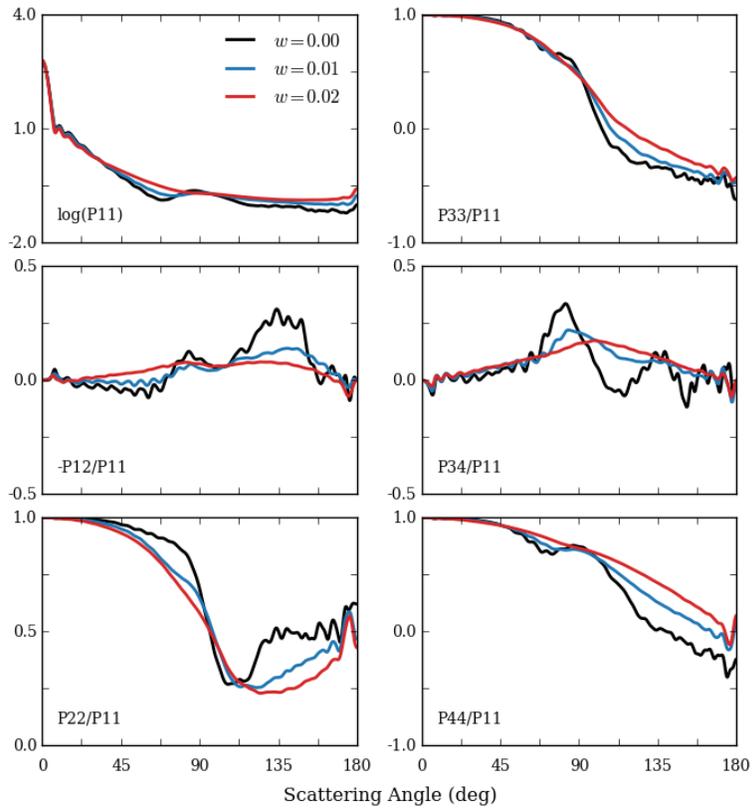}
\caption{The reduced scattering phase matrix elements computed with the II-TM method for spheroidal particles with different surface roughness degrees, size parameter is set to $50$, and $m = 1.5 + 0.001i$.}
\label{fig:iitm}
\end{figure}

For particle with size parameter $X = 60$, it takes about 25 hours wall time  with 100 GB memory usage for a single II-TM simulation. The computing platform used is a IBM NeXtScale nx360 M4 dual socket server based on the Intel Xeon 2.5GHz E5-2670 v2 10-core processor, 2.5 GHz and 256GB RAM, 20 cores on the node. Fig. \ref{fig:iitm} shows the phase matrix elements for randomly rough spheroids with different roughness degree $w$ ( size parameter $X=60$, aspect ratio $\epsilon = a/b = 2$). It shows roughness has little impact on scattering phase function for angles $<5^\circ$. While for angles $>40^\circ$,  roughness will enhance scattering phase function $P_{11}$, especially for the backward direction $180^\circ$, which means backscattering is quite sensitive to surface roughness. For other phase matrix elements like $P_{22}/P_{11}$,surface roughness changes the curve dramatically, all local extrema are smoothed out. Generally, increasing surface roughness will make the curve smoother. Increasing surface roughness, variance of the curves of phase matrix elements decreases. 

\begin{figure}
\centering
\includegraphics[width=1.0\linewidth]{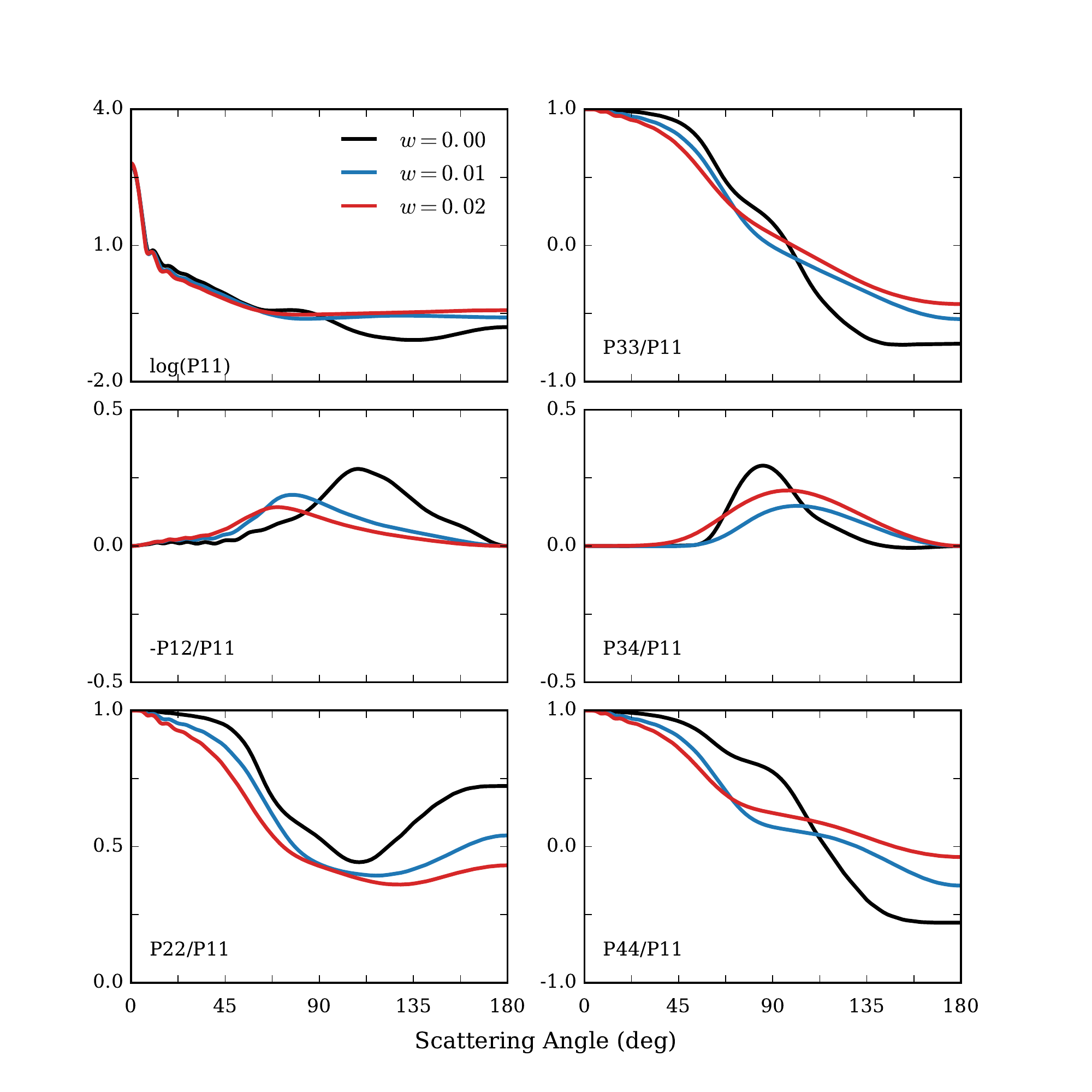}
\caption{The reduced scattering phase matrix elements computed with the IGOM method for spheroidal particles with different surface roughness degrees, size parameter is set to $50$, and $m = 1.5 + 0.001i$. }
\label{fig:igom}
\end{figure}

Fig. \ref{fig:igom} shows phase matrix elements of the same sample particle as above with IGOM.The figure shows consistent moving trend for these two methods. 

We examine the performance of randomly rough spheroids by comparing modeling results with the experimentally measured data for feldspar ensemble\cite{munoz:2012}. To obtain model scattering phase matrix $P(\mu)$, we compute its ensemble average with
\begin{equation}
P(\mu) = \frac{\sum_{s, X}n_s n_X \sigma_{sc}(s, X) P(\mu,s, X)}{\sum_{s, X}  n_s n_X \sigma_{sc}(s, X)}
\end{equation}
where $X$ is the particle size parameter, $s$ represents its shape, $n_X, n_s$ are particle size and shape probability mass function and $\sigma(s, X)$  is the scattering cross section for a particle with shape $s$ and size parameter $X$.  Model particle size range was set according to the measured data \cite{munoz:2012}. Furthermore, we assumed that spheroid shape (parameterized with aspect ratio $\epsilon$, $z = \epsilon$ if $\epsilon > 1$, and $z = -1/\epsilon$ if $\epsilon < 1$) follows a power distribution $p(z) = C|z|^3$(C is the normalization constant.), which was also used in \cite{merikallio:2011}. 

\begin{figure}
\centering
\includegraphics[width=1.0\textwidth]{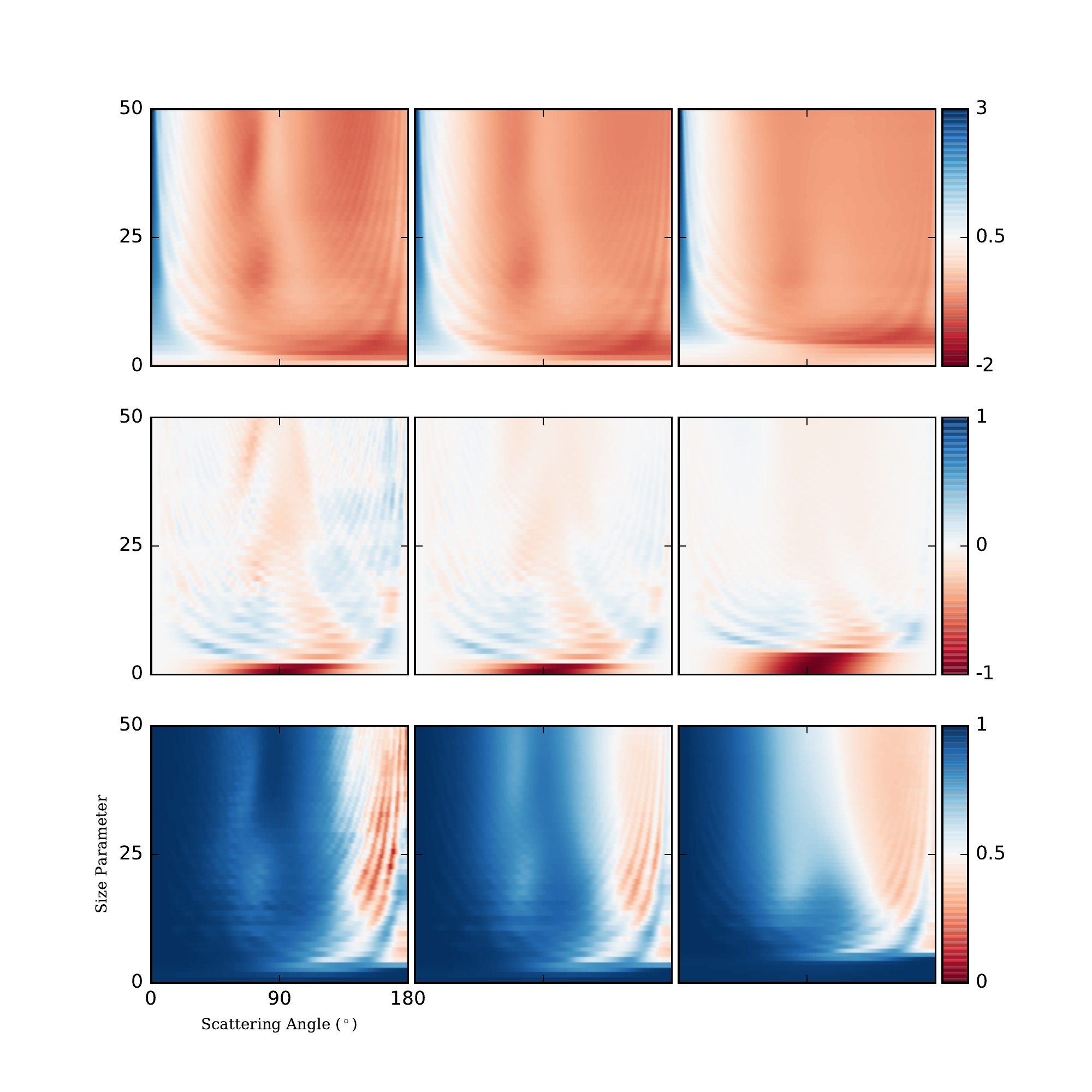}
\caption{Density plots of oblate($b/a= 2.0$) ensemble phase matrix elements (top to bottom: $logP_{11}$, $P_{12}/P_{11}$, $P_{22}/P_{11}$) with different surface height deviation $w$ from left to: 0.00, 0.01, 0.02.}
\label{fig:woblate}
\end{figure}
\begin{figure}
\centering
\includegraphics[width=1.0\textwidth]{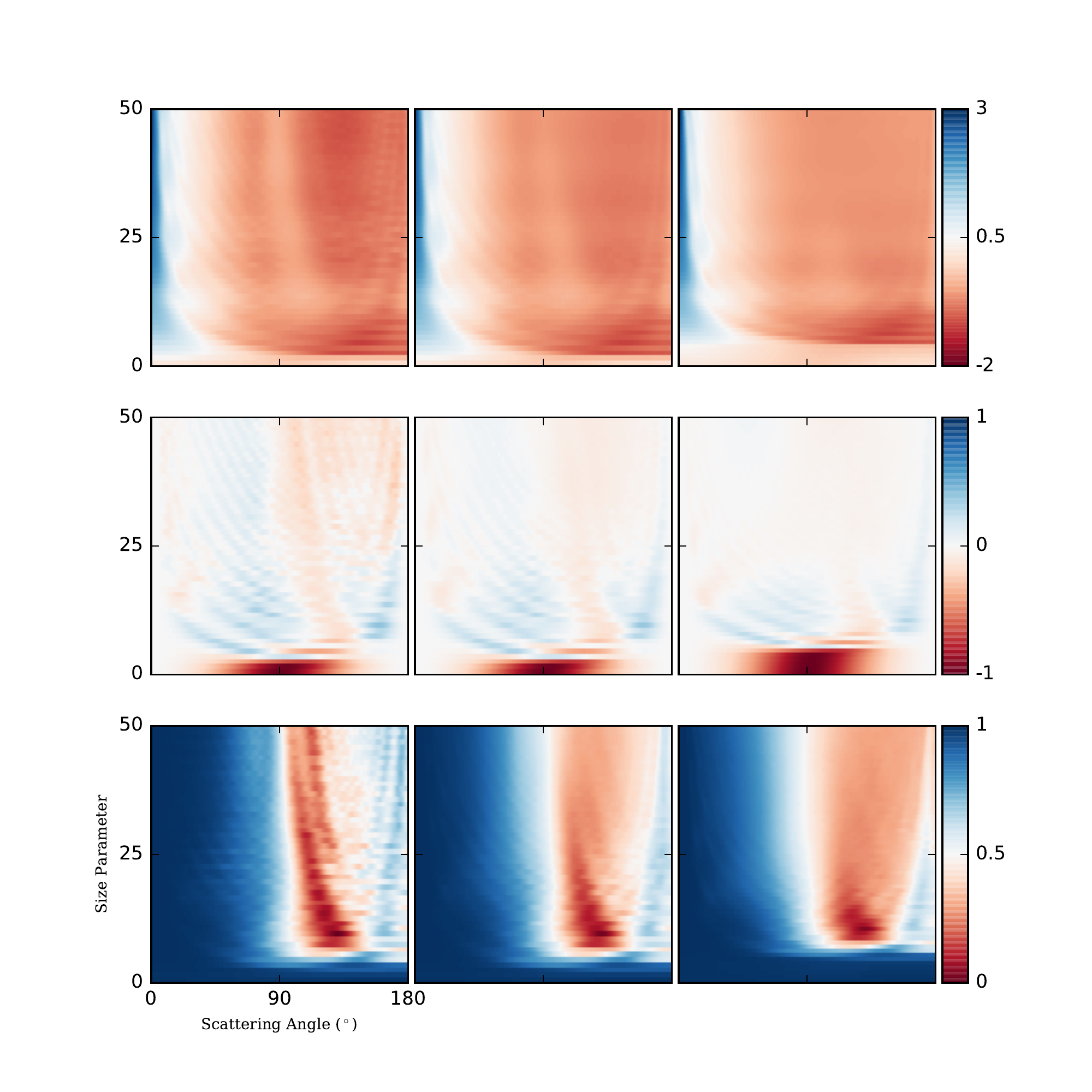}
\caption{Density plots of prolate($a/b = 2.0$) ensemble phase matrix elements (top to bottom: $logP_{11}$, $P_{12}/P_{11}$, $P_{22}/P_{11}$) with different surface height deviation $w$ from left to: 0.00, 0.01, 0.02.}
\label{fig:wprolate}
\end{figure}
\begin{figure}
\centering
\includegraphics[width=1.0\linewidth]{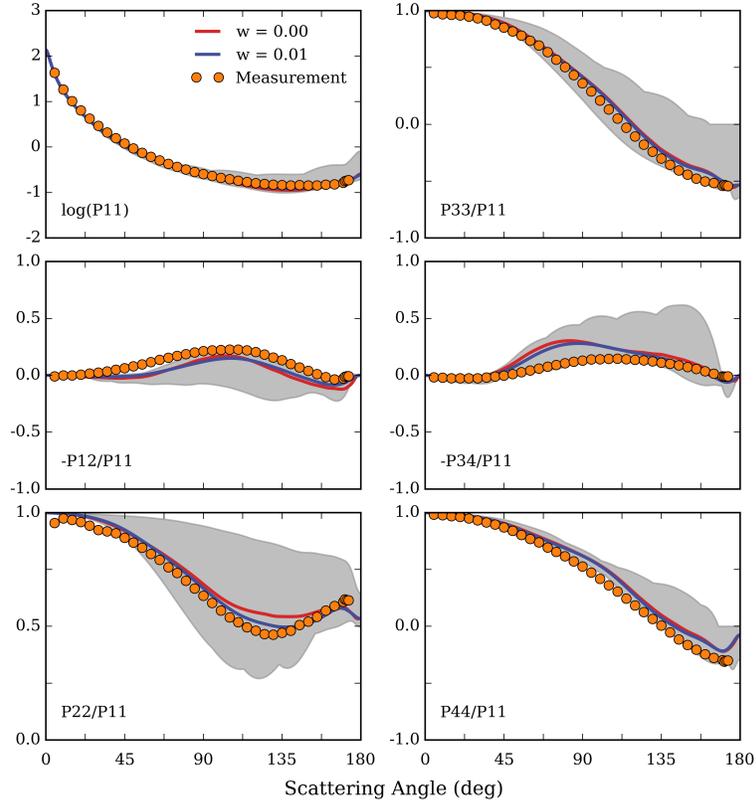}
\caption{Comparison of measured data (orange dots) and modeled scattering phase matrix elements for the same size distribution with different surface roughness degrees at wave-length 632.8 nm. $w = 0.0$ corresponds to the smooth spheroid. $m = 1.5+0.001i$. The shaded area illustrates value coverage of scattering matrix elements by model particles with different size and surface roughness. A power distribution $p(z) = C|z|^n$(C is the normalization constant.) for spheroid shape (aspect ratio $\epsilon$, $z = \epsilon$ if $\epsilon > 1$, and $z = -1/\epsilon$ if $\epsilon < 1$) was used  as \cite{merikallio:2011}, where n was set to be 3. }
\label{fig:measurement}
\end{figure}

Fig.\ref{fig:woblate} and Fig. \ref{fig:wprolate} indicate that the surface roughness has great impact on phase matrix of spheroidal particles, especially larger size. The following comparison with measurements also shows rough model particles have better performance than smooth ones. The optical scattering of dust particles is size dependent. We simulate light scattering by dust aerosols within size parameter ranging from 1 to 300. In the simulation, II-TM is used for size smaller than 60 and IGOM is used for size larger than 60. The estimated refractive index is set to be $m = 1.5 + 0.001i$, and particle size distribution is statistically estimated by observations given in \cite{munoz:2012}. Specifically, we use a volume distribution for dust particles in our comparison. We compare our model results with the measured data from the Amsterdam Light Scattering Database\cite{munoz:2012}. Fig. \ref{fig:measurement} shows model data versus measured data. The model data show good agreement with measurements.  Randomly rough spheroidal particles reproduce most of the phase matrix elements, and give better fit for observed dust scattering data. 

\section{Conclusion}
In conclusion, a method applying discrete differential geometry method to model dust particles' surface roughness is discussed in detail. This method applies a novel approach for generating rough surfaces based on stochastic differential equation and finite element. When applied to particle surface, correlation of surface random field can be estimated. In particular, the correlation length of an random field on particle surface is obtained with a quasi planar approximation.  The mean and variance of the surface random field on each particle can be estimated using classic statistical inference methods. 

Furthermore, information entropy and KL-divergence are introduced for quantitatively distinguishing optical differences between two particles, which can be used as a similarity measure. Application to remote sensing of atmospheric particles and optical characterization of small particles are also possible. Light scattering by rough spheroids are simulated with II-TM and IGOM software package. Finally, ensemble modeling results are compared with the laboratory measured scattering phase matrix data of feldspar dust. 

\section{Acknowledgments}
We would like to thank Ping Yang for helpful discussions and reviewers for their useful suggestions for improvements.  This study was supported by the National Natural Science Foundation of China (NSFC) (Grant No. 41705010) and ``The Fundamental Research Funds for the Central Universities''(DUT16RC(3)121). 

\section*{References}
\bibliography{mybibfile.bib}

\end{document}